\documentclass[prb,showpacs,citeautoscript,floatfix,preprint]{revtex4}
\usepackage{amssymb}
\usepackage{amsmath}
\usepackage{graphicx}
\usepackage{dcolumn}
\usepackage{bm}

\begin{document}

\title{Formal analytical solutions for the Gross-Pitaevskii equation.}
\author{C. Trallero-Giner,$^{(a)}$ Julio C. Drake-Perez,$^{(a)}$ V. L\'{o}%
pez-Richard$^{(b)}$, and Joseph L. Birman$^{(c)}$}
\date{\today }
\affiliation{(a) Faculty of Physics, Havana University, 10400 Havana, Cuba\\
(b) Universidade Federal de S\~{a}o Carlos, Departamento de F\'{\i}sica,
13560-905, S\~{a}o Carlos, SP, Brazil.\\
(c) Department of Physics, The City College of CUNY, New York, NY 10031}

\begin{abstract}
Considering the Gross-Pitaevskii integral equation we are able to formally
obtain an analytical solution for the order parameter $\Phi (x)$ and for the
chemical potential $\mu $ as a function of a unique dimensionless non-linear
parameter $\Lambda $. We report solutions for different range of values for
the repulsive and the attractive non-linear interactions in the condensate.
Also, we study a bright soliton-like variational solution for the order
parameter for positive and negative values of $\Lambda $. Introducing an
accumulated error function we have performed a quantitative analysis with
other well-established methods as: the perturbation theory, the Thomas-Fermi
approximation, and the numerical solution. This study gives a very useful
result establishing the universal range of the $\Lambda $-values where each
solution can be easily implemented. In particular we showed that for $%
\Lambda <-9$, the bright soliton function reproduces the exact solution of
GPE wave function.
\end{abstract}

\pacs{03.75.Be, 03.75.Lm, 05.45.Yv, 05.45.--a}
\maketitle


\section{Introduction}

Since the unambiguous experimental realization in dilute ultra-cold atom
cloud of the Bose-Einstein condensed phase (BEC)\cite%
{Exp1,Exp2,Exp3,Exp4,Exp5,Exp6,Exp7} a lot of work has been devoted for
searching the dynamic and physical properties of the nonlinear matter waves
and excitations of the condensate. The observation of this effect in dilute
atomic gases has allowed to invoke the weakly interacting mean field theory
to describe the properties of the BEC systems.\cite{Groos1,pitaevskii}
Hence, the dynamics of the process for the order parameter has been ruled by
equations of nonlinear Schr\"{o}dinger (NLS) type and mainly by the
Gross-Pitaevskii equation (GPE). Nowadays, NLS equations with attractive
(negative scattering length\cite{khaykovich}) and repulsive (positive
scattering length\cite{burger}) nonlinear interactions have been reported to
describe experimental observations of different types of wave solitons.\cite%
{Eiermann} \ Most of the theoretical work has been devoted to implement
numerical solutions of the GPE for the order parameter (see Ref. [%
\onlinecite{pitaevskii}] and references therein). To study and to control
the physical properties of the condensate it will be very useful to
manipulate analytical expressions for the chemical potential and for the
order parameter as well. A typical example of the great physical interest is
the attention devoted to the collective excitation spectrum of a BEC. In
this case we have to deal with the time-dependent GP equation under the
linear response approximation. Here, as input parameter, we have to insert
in the Bogoliubov equations\cite{Bogoliuv} the order parameter and the
chemical potential solution of the NLS. In that sense, some analytical
results for those magnitudes are priceless. Also, many order problems can be
stressed on the field of cold atoms BEC as the dynamical stability,\cite%
{ruprecht} atomic current in an optical lattice,\cite{Pomoraev} etc.

In order to achieve closed solution the variational procedure with a
Gaussian as a trial wave function has been proposed (see Ref. [%
\onlinecite{1D1}] and reference therein). Nevertheless, it is well know that
this Ansatz does not reproduce well the properties of the condensate. For
example, in the repulsive interaction case and in the strong nonlinear
limit, the shape of the order parameter should be similar to the
Thomas-Fermi (TF) solution (see below Eq. (\ref{TFWF})). Moreover, the
results obtained with the standard variational procedure is, in many cases,
qualitative and, even if the real shape of the wave function resembles the
trial wave function, the variational method is not always a good reference
for solving nonlinear equations.\cite{trallero}

Nowadays, available numerical methods for solving differential equations are
fast and accurate. Nevertheless, if the evaluation of several physical
magnitudes is carried out, such as the optical properties among others, or
to control the properties of the condensate (as we just discussed above),
this advantage is lost due to cumbersome numerical computational procedures
that must be performed at the end of the calculation. Moreover, if we work
with a given basis of functions, it is difficult to know \textit{a priori}
if, in fact, the basis is a complete set for the Hilbert space of the
specific nonlinear equation. Also, the type as well as the swiftness of
convergence to the real solution is not always well established. In that
sense, to implement manageable analytical expressions for the order
parameter, where the accuracy and the absolute error of the obtained
solution are controlled, has becomes a necessity. This is a fact in the
study of nonlinear equation and in particular for the GPE.

A description of the order parameter $\Phi (x)$ in terms of a controlled
truncated basis becomes a useful tool if we are dealing with not many
implemented functions and the degree of accuracy is well established. So,
the obtained expansion will be given by a sum of few basic functions,
allowing in that way to handle with explicit solution to describe the
physical properties of the condensate. Unfortunately, the beauty of such a
mathematical result is restricted to certain range of values for the
parameter involved in the nonlinear equation under study. The challenge is
to find precisely this range of convergence, to give the absolute error in
terms of physical parameters, and to provide other handled compact solutions
outside the obtained range of the desire accuracy. We would like to remark
that the most important requirements for analytical solutions are
simplicity, flexibility, and the viability to be used in perturbation
approaches for the calculations of physical properties.

In this paper we present different methods of solutions of the time
independent GPE based on the equivalent integral GPE\ and its relation with
the Green function of the corresponding linear operator, on the soliton
solution, and on a bright soliton-like variational function. This discussion
will provide general analytical expressions for the order parameter and for
the chemical potential in a universal range of the non-linear interaction
parameter.

To describe the order parameter $\Phi (x)$ we started with the isomorphic
one-dimensional nonlinear Gross-Pitaevskii equation (GPE),\cite%
{Groos1,pitaevskii} which can be written as

\begin{equation}
-\frac{\hbar ^{2}}{2m}\frac{d^{2}\Phi }{dx^{2}}+\frac{1}{2}m\omega
^{2}x^{2}\Phi +\lambda \left\vert \Phi \right\vert ^{2}\Phi =\mu \Phi
\label{GP1}
\end{equation}%
with the normalization condition

\begin{equation}
1=\int dx\left\vert \Phi \right\vert ^{2}.
\end{equation}%
In the above equation $\mu $ represents the chemical potential, $\omega $ is
the trap oscillator frequency, $m$ is the alkaline atom mass, and $\lambda $
is a self-interaction parameter describing the interaction between the
particles$.$

Equation (\ref{GP1}) presents an explicit solution if the non-linear term $%
\left\langle \lambda \left\vert \Phi \right\vert ^{2}\right\rangle $ is
larger than the mean value of kinetic energy operator. This approximation,
known as TF,\cite{edwards,baym} provides simple expressions for the chemical
potential and the wave function given by

\begin{equation}
\frac{\mu _{TF}}{\hbar \omega }=(\frac{3\sqrt{2}}{8}\Lambda )^{2/3},
\label{mui}
\end{equation}%
\begin{equation}
l_{0}\left\vert \Phi _{TF}\right\vert ^{2}=\frac{1}{\Lambda }\left[ (\frac{3%
\sqrt{2}}{8}\Lambda )^{2/3}-\frac{1}{2}(\frac{x}{l_{0}})^{2}\right] ,
\label{TFWF}
\end{equation}%
where $\Lambda =\lambda /l_{0}\hbar \omega ,$ $l_{0}=\sqrt{\hbar /m\omega },$
and the value of $\lambda \geq 0$ is restricted by Eq. (\ref{TFWF}).

The next section is devoted to develop the proposed methods to solve Eq. (%
\ref{GP1}), beyond the above typical TF approximation, . The main goal is to
obtain explicit representations for the whole range of the self-interaction
parameter $\lambda $ (negative and positive values) and to show the range of
validity for each particular method of solution.

\section{Analytical approaches}

First we will study the variational method based on a soliton wave function
as Ansatz function, secondly we analyze the validity of the spectral method
based on the equivalency between the integral and differential equation (\ref%
{GP1}) and the Green function, solution of the linear harmonic oscillator
operator. Moreover, using the obtained general formalism we report
perturbation solutions for $\Phi $ and $\mu $ in terms of the non-linear
parameter $\lambda $. For sake of comparison and in order to check the
accuracy of the implemented approaches, the numerical solution of Eq. (\ref%
{GP1}) is also addressed.

\subsection{Variational method: Soliton approach}

The variational method, valid for positive as well as negative values of $%
\lambda $, could provide a simple picture of the main physical
characteristics of the BEC. Without the trap potential, the GPE (\ref{GP1})
reduces to the nonlinear Schr\"{o}dinger equation which for $\lambda <0$
admits the stationary normalized \textit{bright soliton} solution

\begin{equation}
\Phi _{S}(x,K)=\left( \frac{K}{2}\right) ^{1/2}sech(Kx).  \label{so}
\end{equation}%
Here, the chemical potential $\mu _{S}$ and the inverse of the soliton
length $K$ are expressed by

\begin{equation}
\mu _{S}=-\frac{m\lambda ^{2}}{8\hbar ^{2}},\text{ \ \ \ \ \ \ \ \ \ \ \ }K=%
\frac{m\lambda }{2\hbar ^{2}}.  \label{soExa}
\end{equation}%
In order to solve Eq. (\ref{GP1}) for all values of $\lambda $ we propose as
variational Ansatz the bright soliton (\ref{so}) where $K$ is taken as a
variational parameter. The Ritz's variational method applied to the NLS (\ref%
{GP1}) provides for the chemical potential $\mu (K)$ the parametric equation
(see Appendix A)

\begin{equation}
\mu _{var}(K)=\frac{\hbar ^{2}K^{2}}{2m}\alpha +\frac{m\omega ^{2}}{2K^{2}}%
\beta +\frac{K\lambda }{4}\gamma ,  \label{chva}
\end{equation}%
where $K$ must fulfill the dimensionless equation

\begin{equation}
b^{4}+b^{3}-\delta =0  \label{b}
\end{equation}%
with $b=K\hbar ^{2}/(m\lambda )$ and $\delta =(1/\Lambda )^{4}\pi ^{2}/4$.
Accordingly, Eq. (\ref{chva}) is reduced to the simple relation

\begin{equation}
\frac{\mu _{var}}{\hbar \omega }=-\frac{1}{6}\Lambda ^{2}(b^{2}-\frac{%
3\delta }{b^{2}})  \label{Var}
\end{equation}%
and for the order parameter we get

\begin{equation}
\sqrt{l_{0}}\Phi _{var}=\left( \frac{\Lambda b}{2}\right) ^{1/2}sech(\Lambda
b\frac{x}{l_{0}}).  \label{VFun}
\end{equation}%
Equation (\ref{b}) is a fourth-degree algebraic equation with only one real
physical meaningful solution, which depends on the sign of the non-linear
interaction parameter $\Lambda $. In order to get a more clear view of the
solution for Eq. (\ref{b}), we carry out separate calculations at $\Lambda
=0 $ and for the strong repulsive (attractive) limit $\Lambda \rightarrow
\infty $ ($\Lambda \rightarrow -\infty $).

\begin{figure}[tb]
\includegraphics[width=75mm]{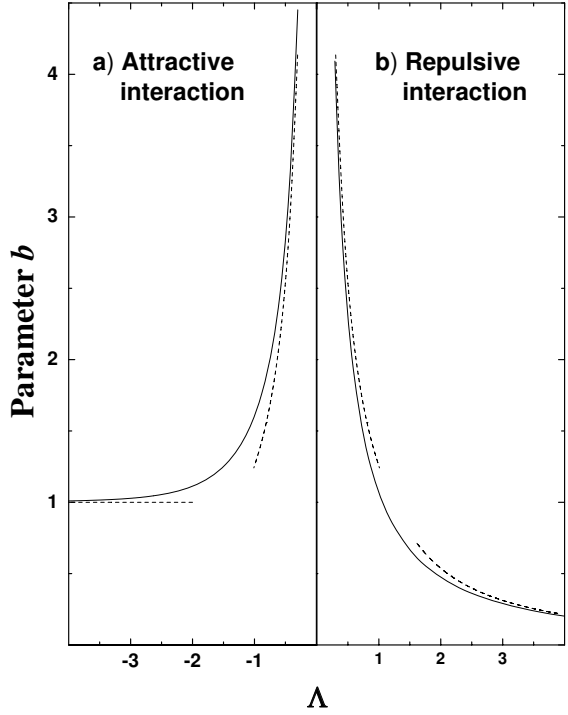}
\caption{Variational parameter $b$ as a function of $\Lambda $. The
asymptotic limits (\protect\ref{bmin}) and (\protect\ref{bmain}),
and the behavior at $\Lambda \approx 0$, Eq. (\protect\ref{b0}), are
indicated by dashed lines.} \label{fig1}
\end{figure}
If $\Lambda \rightarrow 0$ we obtain from Eqs. (\ref{b}) and (\ref{Var})

\begin{equation}
b\times \Lambda =\sqrt{\frac{\pi }{2}}  \label{b0}
\end{equation}%
and%
\begin{equation}
\frac{\mu _{var}(\Lambda =0)}{\hbar \omega }=\frac{\pi }{6}.  \label{lan0}
\end{equation}%
In the strong attractive limit ($\Lambda \ll -1$) and keeping the leading
term in Eq. (\ref{b}), the possible physical solution has the asymptotic
behavior

\begin{equation}
b(\Lambda \rightarrow -\infty )=1+o\left( \sqrt{\frac{\pi }{2}}\frac{1}{%
\Lambda }\right) ^{\frac{7}{3}}  \label{bmin}
\end{equation}%
and the chemical potential $\mu $ is given by%
\begin{equation}
\frac{\mu _{var}(\Lambda \rightarrow -\infty )}{\hbar \omega }=-\frac{1}{6}%
\Lambda ^{2}.
\end{equation}%
In the repulsive limit case, $\Lambda \gg 1,$ Eq. (\ref{b}) yields

\begin{equation}
b(\Lambda \rightarrow \infty )=o\left( \sqrt{\frac{\pi }{2}}\frac{1}{\Lambda
}\right) ^{\frac{4}{3}}+o\left( \sqrt{\frac{\pi }{2}}\frac{1}{\Lambda }%
\right) ^{\frac{7}{3}}  \label{bmain}
\end{equation}%
with%
\begin{equation}
\frac{\mu _{var}(\Lambda \rightarrow \infty )}{\hbar \omega }=\left( \frac{%
\pi \sqrt{2}}{8}\right) ^{\frac{2}{3}}\left( \Lambda \right) ^{\frac{2}{3}}.
\end{equation}%
We can compare the above limit solutions with those obtained by the TF
approximation, Eq. (\ref{mui}), and the exact soliton solution, Eq. (\ref%
{soExa}). The relative errors are equal to

\begin{equation}
\left\vert \frac{\mu _{var}(\infty )-\mu _{TF}(\infty )}{\mu _{TF}(\infty )}%
\right\vert =\left( \frac{\pi }{3}\right) ^{\frac{2}{3}}-1\approx 0.0312.
\label{errorTF}
\end{equation}%
and

\begin{equation}
\left\vert \frac{\mu _{var}(-\infty )-\mu s(\infty )}{\mu s(\infty )}%
\right\vert =\left\vert -\frac{8}{6}+1\right\vert \approx 0.3333.
\end{equation}%
From the above relations we conclude that the variational wave function (\ref%
{so}) provides a better solution of the GPE in the strong repulsive case
than for the attractive one. At $\Lambda =0$ a relative error of 0.0472 is
reached by comparing Eq. (\ref{lan0}) with the exact solution of the
harmonic oscillator problem $\mu /\hbar \omega =0.5$. In Fig. 1 we present
the parameter $b,$ solution of the Eq. (\ref{b}), as a function of the
dimensionless parameter $\Lambda $. Also, the limit solutions are indicated
by dashed lines. It can be seen that the calculated asymptotic behaviors (%
\ref{b0}), (\ref{bmin}), and (\ref{bmain}) at $\Lambda =0,$ $\Lambda
\rightarrow -\infty ,$ and $\Lambda \rightarrow \infty ,$ respectively, are
quickly reached by the exact solutions of Eq. (\ref{b}). It is not
surprising that the Ritz's variational method failed to get a closed
analytical solution of the differential GPE. The variational method here
implemented is only valid for linear differential equations or the
corresponding Lagrangian of the problem.

\subsection{GP integral equation: Green function solution}

One of the most powerful analytical method used to solve differential and
integral equations corresponds to the Green function formalism (GFF). In
order to implement this mathematical technique to the non-linear Schr\"{o}%
dinger equation we rewrite (\ref{GP1}) as
\begin{equation}
L_{0}\left[ \Phi \right] =-\frac{\hbar ^{2}}{2m}\frac{d^{2}\Phi }{dx^{2}}+%
\frac{1}{2}m\omega ^{2}x^{2}\Phi =f(x).  \label{green}
\end{equation}%
Here $f(x)$ will be considered as an inhomogeneity in the differential
equation and equal to
\begin{equation*}
f(x)=(\mu -\lambda \left\vert \Phi (x)\right\vert ^{2})\Phi (x).
\end{equation*}%
Function $\Phi (x),$ solution of Eq. (\ref{green}), can be cast in terms of
the Green function $G(x,x^{\prime })$ of the linear operator $L_{0}\left[
\Phi \right] .$ Formally, we can write $\Phi (x)$ as a function of the
inhomogeneity $f(x)$ as\cite{morse}

\begin{equation}
\Phi (x)=\int_{-\infty }^{\infty }G(x,x^{\prime })(\mu -\lambda \left\vert
\Phi (x^{\prime })\right\vert ^{2})\Phi (x^{\prime })dx^{\prime }.
\label{in}
\end{equation}%
The above expression corresponds to the GP integral equation for the order
parameter $\Phi (x)$ \ We observe that the integral equation (\ref{in}) has
a symmetric kernel, $G(x,x^{\prime }),$ which fulfills the differential
equation
\begin{equation*}
L_{0}\left[ G(x,x^{\prime })\right] =\delta (x-x^{\prime }).
\end{equation*}%
To write the formal solution (\ref{in}) in terms of the Green function of
the operator $L_{0}$, the function $f(x)$ has some constrains.\cite%
{Mihling,petrovskii} In our case, all functions and the Green function also,
have to fulfill the boundary condition $\Phi (x)\rightarrow 0$ as $%
x\rightarrow \pm \infty .$. This guarantees that the inhomogeneity $f(x)$
belongs to the same Hilbert space of the linear operator $L_{0}.$ The kernel
$G(x,x^{\prime })$ is the given by the following spectral representation

\begin{equation}
G(x,x^{\prime })=\sum_{n=0}^{\infty }\frac{\varphi _{n}(x)\varphi
_{n}(x^{\prime })}{\hbar \omega (n+1/2)},  \label{gf}
\end{equation}%
with $\varphi _{n}(x)$ being the harmonic oscillator wave function\cite%
{Abramowitz}

\begin{equation}
\varphi _{n}(x)=\left( \frac{1}{\pi ^{1/2}2^{n}n!l_{0}}\right) ^{1/2}\exp
\left( \frac{-x^{2}}{2l_{0}{}^{2}}\right) H_{n}\left( \frac{x}{l_{0}}\right)
.  \label{OSC}
\end{equation}

We have to note that according to the general theory of Fredholm integral
equations\cite{Mihling,petrovskii}, the set of functions appearing in the
spectral representation of a symmetric kernel, $\{\varphi _{n}(x)\}$ in the
present case, represents a complete set of functions for the given Hilbert
space of the GP integral equation (\ref{in}). Hence, the convergence of the
expansion (\ref{gf}) is guaranteed and we can insert the spectral
representation of $G(x,x^{\prime })$ in (\ref{in}) and interchange the
integral and infinity expansion (\ref{gf}). Thus

\begin{equation}
\Phi =\sum_{n=0}^{\infty }\frac{\varphi _{n}(x)}{\hbar \omega (n+1/2)}\int
\varphi _{n}(x^{\prime })(\mu -\lambda \left\vert \Phi (x^{\prime
})\right\vert ^{2})\Phi (x^{\prime })dx^{\prime }.  \label{gf3}
\end{equation}%
From (\ref{gf3}) it is straightforward that the general solution for the
order parameter $\Phi $ has an explicit representation through the harmonic
oscillator $\varphi _{n}(x)$ as
\begin{equation}
\sum_{n=0}^{\infty }\varphi _{n}(x)C_{n}(\mu ).  \label{fi}
\end{equation}%
Since the inhomogeneity $f(x)$ belongs to the same Hilbert space of the
symmetric kernel of the of Fredholm integral equation (\ref{in}), the
convergency of the series (\ref{fi}) in energy to the function $\Phi $ is
guaranteed.\cite{MihlingII}

In the present case the coefficients $C_{n}(\mu )$ are restricted to obey
the relation

\begin{equation}
C_{n}=\int \frac{1}{\hbar \omega (n+1/2)}\varphi _{n}(x^{\prime })(\mu
-\lambda \left\vert \Phi (x^{\prime })\right\vert ^{2})\Phi (x^{\prime
})dx^{\prime }.  \label{13}
\end{equation}%
Inserting the convergent series (\ref{fi}) in Eq. (\ref{13}), it follows
that the vector coefficient $\mathbf{C}(\mu )$ must fulfil the non-linear
equation system

\begin{equation}
\left[ \mathbf{\Delta }(\mu )+\Lambda \overline{\mathbf{C}}\mathbf{\cdot T}%
\cdot \mathbf{C}\right] \mathbf{C}=0,  \label{Hill}
\end{equation}%
where

\begin{equation}
\mathbf{\Delta }_{nm}=\left( n+\frac{1}{2}-\frac{\mu }{\hbar \omega }\right)
\delta _{nm}\text{.}  \label{matriele}
\end{equation}%
and $T_{plmn}$ is a fourth dimensional matrix defined in the Appendix B.

The order parameter $\Phi $ and the chemical potential as a function of the
dimensionless parameter $\Lambda $ are obtained by solving the non-linear
equation system (\ref{Hill}). Although the mathematical complexity of Eq. (%
\ref{GP1}) has been reduced, Eq. (\ref{Hill}) is nevertheless an infinite
generalized eigenvalue problem for $\mu (\Lambda )$ and $\mathbf{C}(\mu
(\Lambda ))$. The complexity of the problem depends on the sign and the
values of the non-lineal parameter $\lambda $ but the key issue is how
quickly converges the series in (\ref{fi}) or equivalently, the non-linear
equation system (\ref{Hill}). This important problem is addressed in the
next section.

We have to mention that the obtained problem (\ref{Hill}) is isomorphic to
Galerkin method. The former one is a generalized variational method where
for a given equation $L[F]=L_{0}[F]+L_{p}[F]$ it is possible to choose a
certain basis $\{g_{k}\}$ of the operator $L_{0}$ and to expand the function
$F$ in term of the given basis. The choice of the operator $L_{0}$ (which
must include the boundary conditions) is not unique and certain degree of
freedom prevails. To guarantee that the expansion converge to the real
solution, the picked out operators $L_{0}$ and $L_{p}$ have to fulfil
certain mathematical conditions (see Ref. [\onlinecite{MihlingII}] for a
detailed description of this mathematical treatment). This crucial question
is not trivial when we are dealing with non-linear equations as the NLS. In
our case the mathematical treatment above developed is based on the
properties of the Fredholm integral equations and can be considered a
rigorous demonstration of the validity of the expansion (\ref{fi}) and the
convergence to the correct solution.

We obtained the ground state solution $\Phi _{0}$ in terms of a truncated
basis set, $\{\varphi _{n}(x)\}$ $(n=1,...I)$, by defining the finite
dimensional nonlinear Hill determinant eigenvalue equation \cite{Hill}

\begin{equation}
\left\Vert \mathcal{M}^{(I)}(\mu (\Lambda ),\mathbf{C})\right\Vert =0,
\label{matriz}
\end{equation}%
where $\mathcal{M}_{nm}^{(I)}$, $(n,m=1,2,...I)$ are the corresponding
matrix elements according to the Eq. (\ref{Hill}). Since the scaling of any
direct numerical algorithm of integration implemented to obtain the tensor $%
T_{plmn}$ is of the other $I^{4}\times P$ ($P$ is the number of grid points)
the numerical implementation becomes a cumbersome task and non-efficient
method of evaluation. To get a better efficient algorithm than those based
on a direct numerical integration of the tensor $T_{plmn},$ it is necessary
to exploit its analytical representation together with its symmetry
properties. This analysis is presented in the Appendix B allowing a
straightforward evaluation of the tensor $T_{plmn}$.

To solve Eq. (\ref{matriz}), we have implemented the Neumann iterative
procedure in a finite basis of dimension $I.$ For a given iteration and
since the functions $\{\varphi _{n}(x)\}$ define a complete set for the GPE,
obeying the natural boundary conditions, $\varphi _{n}(x)\rightarrow 0$ for x%
$\rightarrow \pm \infty ,$ the roots of the determinant (\ref{matriz})
converge to the exact ground state solution of Eq. (\ref{in}) and $%
\lim_{I\rightarrow \infty }\mu ^{(I)}(\Lambda )=\mu (\Lambda ).$\cite{Hill}
The numerical procedure starts from a trial vector $\mathbf{C}\backsim $ $%
\widetilde{\mathbf{C}}$ and iteratively we obtain the \ $k-th$
approximation. In each step, the matrix (\ref{matriz}) must be recalculated
by using the new eigenvector $\mathbf{C}\backsim \widetilde{\mathbf{C}}$.
The procedure is repeated until\ \ $\left\vert
C_{n}^{(k)}-C_{n}^{(k-1)}\right\vert <\delta _{c}$ and (or alternatively) $%
\left\vert \mu ^{(k)}-\mu ^{(k-1)}\right\vert <\hbar \omega \cdot \delta
_{\mu },$ where \ $\delta _{c}$\ and \ $\delta _{\mu }$\ \ are the desirable
accuracies for the coefficients and the chemical potential, respectively.
For the iterative procedure, it is useful to introduce a control parameter \
$\varepsilon \in \lbrack 0,1]$, \ so that

\begin{equation*}
\widetilde{C_{n}}=\sqrt{\varepsilon (C_{n}^{(k-1)})^{2}+(1-\varepsilon
)(C_{n}^{(k)})^{2}}.
\end{equation*}%
This procedure is faster and accurate for positive and small negative values
of the non-linear parameter $\Lambda >-5.$ For \ $\Lambda <-5$\ however, the
size of the matrix we have to deal with grows as $\left\vert \Lambda
\right\vert $ does. In the former case, a basis set of 25 functions allows
at least 5 significant figures in the calculation of $\mu $, while for the
later at least 50 oscillator wave functions \{$\varphi _{n}\}$ were sorted
in order to reach the same accuracy at $\Lambda =-10.$

According to the Neumann iterative procedure we have to introduce an initial
starting $\mathbf{C}^{(0)}$ vector. This vector can be chosen according to
the desirable $\Lambda $ value and the following criteria can be
established: i) For the dimensionless interaction parameters $\left\vert
\Lambda \right\vert <1.5,$ the coefficients $C_{n,m}^{(0)}=\delta _{n,m}.$
ii) If $\Lambda >5$ the asymptotic limit of the TF approximation wave
function given by (\ref{TFWF}) is a good starting iterative procedure. iii)
For attractive interaction and $\Lambda <-1.5$ the soliton wave function
approach (\ref{so}) is useful as initial condition .

\subsection{Perturbation theory}

It is useful to get expressions for $\mu $ and the order parameter $\Phi $
through a perturbation approach since these are easily handled solutions.
Also, the explicit perturbation expressions can be implemented as a method
to control other solutions in particular the numerical ones. If the
nonlinear term $H_{p}=\lambda \left\vert \Phi \right\vert ^{2}$ is
considered as a perturbation in comparison to the trap potential $m\omega
^{2}x^{2}/2$, the chemical potential and the vector $\mathbf{C}$ in Eq. (\ref%
{Hill}) can be sought in the form of series, i.e.
\begin{eqnarray*}
C_{m} &=&C_{m}^{(0)}+\lambda C_{m}^{(1)}+\lambda ^{2}C_{m}^{(2)}+...., \\
\mu &=&\mu ^{(0)}+\lambda \mu ^{(1)}+\lambda ^{2}\mu ^{(2)}+......
\end{eqnarray*}%
Taking only the second order interaction in $\lambda ,$ Eq. (\ref{Hill})
yields
\begin{equation}
\mu =\frac{\hbar \omega }{2}+\frac{\lambda }{l_{0}}T_{0000}-3\left( \frac{%
\lambda }{l_{0}}\right) ^{2}\sum_{m=1}^{\infty }\frac{\left\vert
T_{000m}\right\vert ^{2}}{\hbar \omega m}.
\end{equation}%
Using the properties of the matrix $T_{plmn}$ given in the Appendix B we get
\begin{equation}
\frac{\mu }{\hbar \omega }=\frac{1}{2}+\frac{\Lambda }{\sqrt{2\pi }}-\frac{3%
}{2\pi }\Lambda ^{2}\sum_{m=1}^{\infty }\frac{(2m-1)!}{2^{4m}(m!)^{2}}.
\label{pert}
\end{equation}%
Using that

\begin{equation*}
\sum_{m=1}^{\infty }\frac{(2m)!}{2^{3m}(m!)^{2}m(x^{2}+1)^{m}}=-2\ln \left(
\frac{1}{2}\sqrt{\frac{2x^{2}+1}{2(x^{2}+1)}}+\frac{1}{2}\right)
\end{equation*}%
the chemical potential up to second order is reduced to the following useful
expression

\begin{equation}
\frac{\mu }{\hbar \omega }=\frac{1}{2}+\frac{\Lambda }{\sqrt{2\pi }}%
-0.033106\times \Lambda ^{2}.  \label{pertmu}
\end{equation}%
Finally, the normalized order parameter $\Phi $ including terms to the first
order can be expressed as
\begin{equation}
\Phi =\varphi _{_{0}}(x)+\frac{\Lambda }{\sqrt{2\pi }}\sum_{m=1}^{\infty }%
\frac{(-1)^{m+1}\sqrt{(2m)!}}{2^{2m}(m!)2m}\varphi _{2m}(x).  \label{fiper}
\end{equation}

\begin{figure}[tb!]
\includegraphics[width=75mm]{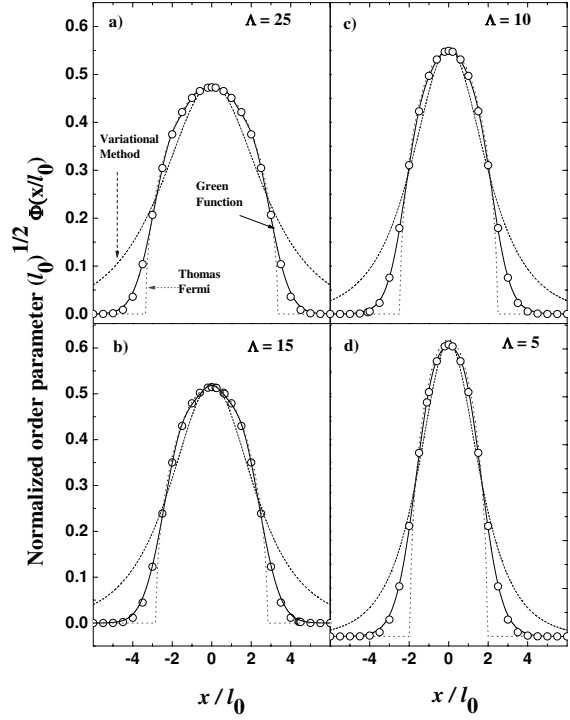}
\caption{(Color online) Normalized order parameter
$\protect\sqrt{l_{0}}\Phi (x/l_{0})$ for the positive dimensionless
self-interaction $\Lambda$ values: a) 25, b) 15, c) 10, and d) 5.
Solid line: Solution (\protect\ref{fi}). Dashed line: Soliton
variational approach. Dot: Thomas-Fermi approximation. Empty
circles: Numerical solution.} \label{fig2}
\end{figure}

\subsection{Numerical Solution}

A comparison of the obtained analytical solutions with direct numerical
calculations is an important control for validating the mathematical methods
here introduced. In order to solve (\ref{GP1}) numerically, we choose a
finite difference method where for the second derivative we select a simple
three-points approximation with uniform spacing, so that the differential
equation can be rewritten as a symmetrical tri-diagonal matrix. The
eigenvalue problem for the obtained matrix can then be solved by the usual
methods. Explicitly we have
\begin{figure}[tb]
\includegraphics[width=65mm]{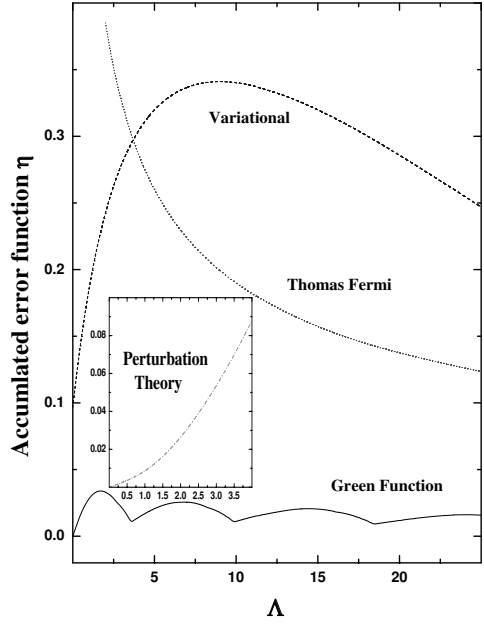}
\caption{(Color online) Accumulated error function $\protect\eta $
for the
repulsive interaction as a function of $\Lambda $. Solution (\protect\ref{fi}%
) (solid line), soliton variational approach (\protect\ref{VFun}) (dashed
line), and Thomas-Fermi function (\protect\ref{TFWF}) (dot line). Inset:
Perturbation wave function (\protect\ref{fiper}) (dot-dashed line).}
\label{fig3}
\end{figure}
\begin{eqnarray}
&&%
\begin{pmatrix}
v_{1} & -\delta ^{-2} &  &  &  &  \\
-\delta ^{-2} & v_{2} & . &  &  &  \\
& . &  & . &  &  \\
&  & . &  & . &  \\
&  &  & . & v_{L-1} & -\delta ^{-2} \\
&  &  &  & -\delta ^{-2} & v_{L}%
\end{pmatrix}%
\begin{pmatrix}
\overline{\Phi _{1}} \\
\overline{\Phi _{2}} \\
. \\
. \\
\overline{\Phi _{L-1}} \\
\overline{\Phi _{L}}%
\end{pmatrix}
\notag \\
&=&\frac{\mu }{\hbar \omega }%
\begin{pmatrix}
\overline{\Phi _{1}} \\
\overline{\Phi _{2}} \\
. \\
. \\
\overline{\Phi _{L-1}} \\
\overline{\Phi _{L}}%
\end{pmatrix}%
,  \label{numeric}
\end{eqnarray}%
where

\begin{eqnarray}
v_{i} &=&v_{i}(\overline{\Phi _{i}})=\frac{1}{2}\left( -\frac{L}{2l_{0}}%
+(i-1)\delta \right) ^{2}  \notag \\
&&+\Lambda \left\vert \overline{\Phi _{i}}\right\vert ^{2}\delta
^{-1}+\delta ^{-2},  \label{nu}
\end{eqnarray}%
$\overline{\Phi _{i}}=\sqrt{l_{0}}\Phi _{i},\ \overline{\Phi _{i}}=\
\overline{\Phi }(-\frac{L}{2l_{0}}+(i-1)\delta ),$ $i=1,2,...L,$ and $\delta
$ is the discreet step. The presence of the wave function inside the matrix
in the left side of Eq. (\ref{numeric}) enforces the use of some kind of
iteration procedure in order to solve the non-linear problem. That is, for
the vector $\mathbf{v}$ of components $v_{i}$ (see Eq. (\ref{nu})) we set

\begin{equation}
\mathbf{v}\left[ \mathbf{F}^{(k)}\right] ,\text{ \ \ \ \ \ \ }k=0,1,2,...,
\label{vectorV}
\end{equation}%
where $\mathbf{F}^{(k)}$ is a certain trial function. We started with
certain $F^{(0)}(x)=\overline{\Phi ^{(0)}(x)}$ evaluated at the $x_{i}$ mesh
points. After that, we find the approximate eigenvector $\Phi $ and
eigenvalue $\mu $ of the ground state solution of Eq. (\ref{numeric}). The
new trial function $\mathbf{F}^{(k)}$ is obtained by the expression

\begin{equation*}
F_{i}^{(k)}=\sqrt{\varepsilon \left[ \overline{\Phi _{i}^{(k-1)}}\right]
^{2}+(1-\varepsilon )\left[ \overline{\Phi _{i}^{(k)})}\right] ^{2}}
\end{equation*}%
with \ $\varepsilon \in \lbrack 0,1].$ This procedure is repeated until $%
\left\vert \overline{\Phi _{i}^{(k)}}-\overline{\Phi _{i}^{(k-1)}}%
\right\vert <\delta _{\Phi }$ and (or alternatively) $\left\vert \mu
^{(k)}-\mu ^{(k-1)}\right\vert <\hbar \omega \cdot \delta _{\mu }$,\ where\ $%
\delta _{\Phi }$\ and $\delta _{\mu }$\ are the desirable accuracies for the
wave function and the chemical potential, respectively. A similar procedure
has been used with success in Ref. [\onlinecite{Pu}] for a two component
BEC. The practical implementation of the above described method is mainly
straightforward, however, due to the influence of the non-linear term, the
accuracy and speed of convergence is critically dependent on the correct
choice of the parameter $\varepsilon $. Our experience shows that the best
value $\varepsilon $ depends on the value and sign of the non-linear term.

\begin{figure}[tb!]
\includegraphics[width=75mm]{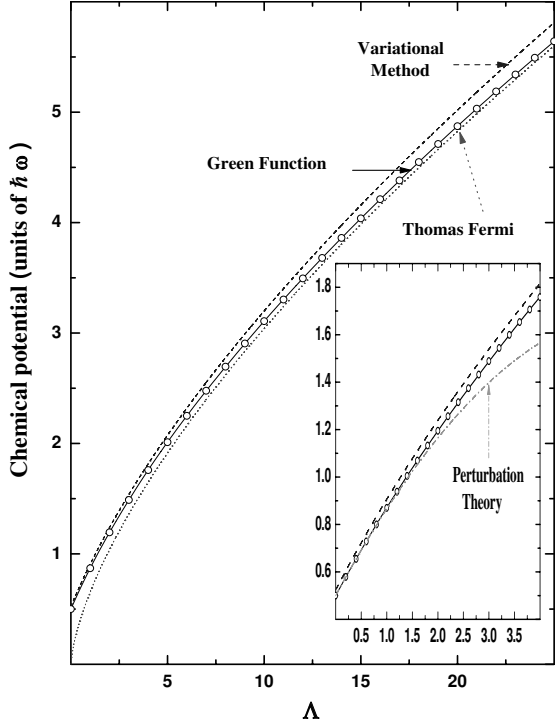}
\caption{(Color online) Chemical potential in units of $\hbar \protect\omega
$ as a function of dimensionless self-interaction parameter $\Lambda $.
Solid line: Equation (\protect\ref{Hill}). Dashed line: Soliton variational
approach (\protect\ref{Var}). Dotted line: Thomas-Fermi approximation (%
\protect\ref{mui}). Empty dot: Numerical solution. Inset: Perturbation
theory (\protect\ref{pertmu}) (dot-dashed line).}
\label{fig4}
\end{figure}

\section{Results}

We shall now discuss the accuracy and the reliability of the above
implemented methods of solution, by studying independently the repulsive and
the attractive interaction cases.
\begin{figure}[tb]
\includegraphics[width=75mm]{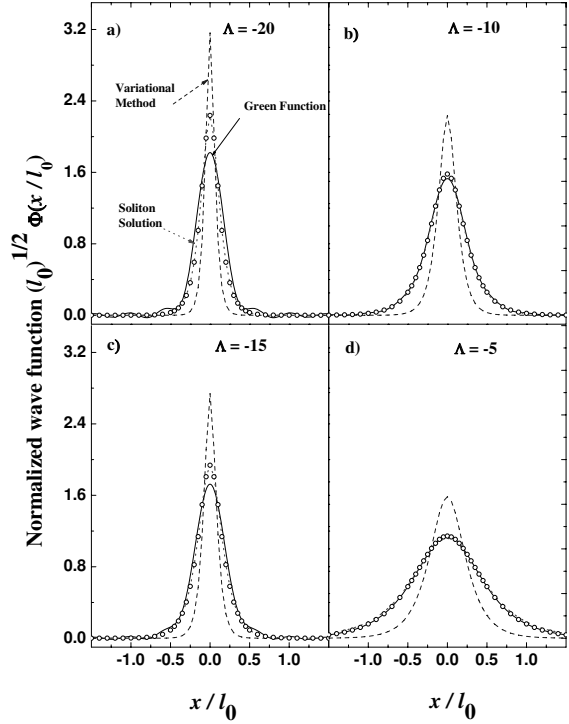}
\caption{(Color online) The same as Fig. \protect\ref{fig2} for the
attractive dimensionless self-interaction $\Lambda $ values: a) -20, b) -15,
c) -10, and d) -5. Dotted line represents the soliton solution (\protect\ref%
{so}).}
\label{fig5}
\end{figure}

\subsection{Repulsive interaction}

Figure 2 displays the order parameter $\sqrt{l_{0}}\Phi (x/l_{0})$ for
several values of $\Lambda $. The variational solution of the Eqs. (\ref{Var}%
) and (\ref{VFun}) is represented by dashed lines, solid lines present the
calculation using Eq. (\ref{fi}), and the TF approach, following Eq. (\ref%
{TFWF}), is indicated by dots. Empty circles show the obtained numerical
solutions of the GPE (\ref{GP1}) following the procedure described in Sec.
II D. It can be seen that the wave function is more delocalized and the
maximum of $\Phi (x/l_{0})$ decreases as $\Lambda $ increases, thus the
condensate spreads as the non-linear term increases. Also, in Fig. 2, the
differences between all calculated analytical representations and the
numerical procedure are qualitatively displayed.\ As already known, the TF
approach reproduces well the properties of the condensate for large values
of $\Lambda $, while the proposed variational solution exhibits a better
approximation for small values of $\Lambda $. In general, we have obtained
very good agreement between the solution (\ref{fi}) and the numerical
solution for all considered values of dimensionless interaction parameter $%
\Lambda .$ Nevertheless, it is useful to define a magnitude that quantify
the quality of the implemented analytical solutions. Hence, we have
introduced the accumulated error function%
\begin{equation}
\eta _{i}=\int_{-\infty }^{\infty }\left\vert \Phi _{num}(x)-\Phi
_{i}(x)\right\vert dx,  \label{error}
\end{equation}%
where $\Phi _{num}$ is the numerical solution of Eq. (1). The above
magnitude gives a direct estimation of the total error introduced throughout
the whole interval $-\infty <x<\infty $. Since in each given point $x\in
(-\infty ,\infty )$ we add the modulus of the difference between $\Phi
_{num}(x)$ and $\Phi _{i}(x)$, then $\eta _{i}$ determines the maximum
accumulated error for the analytical wave function $\Phi _{i}(x).$ Figure 3
presents the estimated error $\eta _{i}$ as a function of the dimensionless
interaction term $\Lambda $ for all functions considered: the Thomas-Fermi
(dot line), the solution (\ref{fi}) (solid line), and the soliton
variational approach (dash line). From the figure it can be seen that the
best analytical solution ($\eta <0.033$) is reached by using (\ref{fi}),
while the TF approximation approaches, asymptotically, to the exact solution
. The soliton variational solution exhibits its minimum error ($\eta <0.2$)
for $\Lambda <2$, reaching a maximum error at $\Lambda \approx 10.$ One can
notice that $\Phi _{var}$ is a better approach than the TF for $\Lambda <3.6.
$ The accumulative error introduced by the perturbation wave function
(dash-dot line) is also shown in the inset. In general, the accuracy of the
series (\ref{fi}) can be greatly improved if large matrixes are implemented.
In our calculations a few functions (a $50\times 50$ matrix) was necessary
to achieve an accuracy of $10^{-8}$ for the chemical potential. In the case
of the numerical procedure (\ref{numeric}), values of $\mu /\hbar \omega $
were calculated with an uncertainty of $10^{-10}$.

\begin{figure}[tb]
\includegraphics[width=65mm]{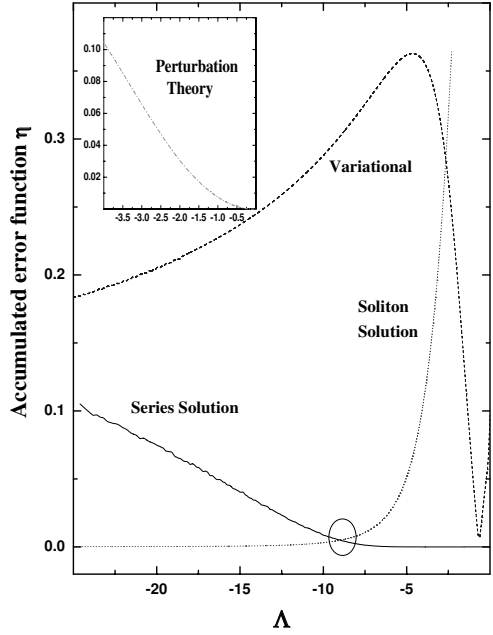}
\caption{(Color online) The sane as Fig. \protect\ref{fig3} for the
attractive interaction as a function of $\Lambda$. Dotted line represents
the soliton solution (\protect\ref{so}). Inset: Accumulated error function $%
\protect\eta $ for the perturbation wave function (\protect\ref{fiper})
(dot-dashed line).}
\label{fig6}
\end{figure}

In Fig. 4, we compare the calculated chemical potential in units of the
energy trap $\hbar \omega $ according to the analytical methods outlined in
the previous section. The Thomas-Fermi approach following Eq. (\ref{mui}) is
indicated by dots, the soliton variational solution obtained by solving the
Eqs. (\ref{b}) and (\ref{Var}) is represented by a dashed line, while the
solid line presents the calculation using the Hill determinant (\ref{matriz}%
). In the inset, the comparison with the perturbation theory given by Eq. (%
\ref{pertmu}) (dash dot line) is also shown. The numerical solution is also
presented by empty dots. As expected the TF limit increases its accuracy,
i.e., less than 3\% of error at $\Lambda =10,$ as the non-linear parameter
increases. No differences can be observed in the scale of the figure between
the numerical solution and the chemical potential using Eq. (\ref{matriz}).
The soliton variational calculation presents a larger error for $\Lambda >7.$
Also, in the figure we can observe the relative error of $0.0312$ between
the TF and variational solutions as reported by the Eq. (\ref{errorTF}) at $%
\Lambda \rightarrow \infty $. Concerning the perturbation theory, the best
accuracy, lees than $3\%,$ is reached for $\Lambda <2.$ The results shown in
the Figs. 3 and 4 have a universal character and the comparison between the
analytical methods provides universal criteria of their validity ranges.

\subsection{Attractive interaction}

Following the same trends as in the repulsive case, Fig. 5 shows the
normalized order parameter for four negative values of $\Lambda $. In the
figure, the dotted line represents the soliton solution (\ref{so}). We
observe that using the series (\ref{fi}) the agreement is not so wide
ranging as for the repulsive case . For values of $\Lambda >-10,$ we obtain
a better match between the Eq. (\ref{fi}) and the numerical solution (\ref%
{numeric}). Nevertheless, the agreement reached
\begin{figure}[b]
\includegraphics[width=75mm]{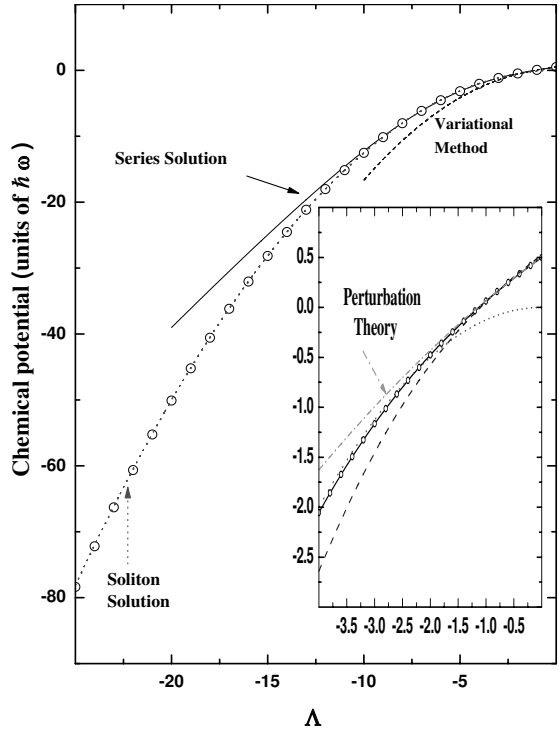}
\caption{(Color online) The same as Fig. \protect\ref{fig4} for the
attractive interaction as a function of $\Lambda $. Dotted line represents
the soliton chemical potential (\protect\ref{soExa}). Inset: Perturbation
theory (\protect\ref{pertmu}) (dot-dashed line) }
\label{fig7}
\end{figure}
\begin{figure}[tb]
\includegraphics[width=75mm]{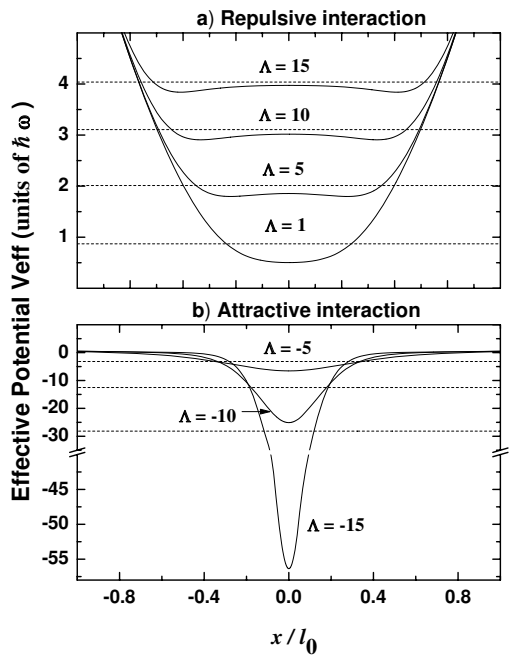}
\caption{Effective Potential $Veff$ for the GPE (see text).}
\label{fig8}
\end{figure}
with the soliton solution (\ref{so}) is remarkably good. In order to
quantify the discrepancy between the implemented analytical solutions and
the numerical one, we evaluate the accumulated error (\ref{error}) in terms
of $\Lambda $. Figure 6 presents $\eta _{i}$ for all considered functions $%
\Phi _{i}.$ Here, a dotted line is used for the soliton solution (\ref{so}).
We have estimated that the best result by using Eq. (\ref{fi}) is reached,
for $\Lambda \gtrsim -10$, while the exact soliton solution gives a better
approach for $\Lambda <-10,$ and in both cases we have an accumulated error $%
\eta <0.005$. The soliton variational solution $\Phi _{var}$ yields a
maximum error of $\eta \approx 0.36$ at $\Lambda \approx -4.6.$ The
accumulated error using the perturbation wave function (dash-dot line) is
also shown in the inset and as expected $\eta \rightarrow 0$ as $\Lambda
\rightarrow 0$.

Figure 7 depicts the calculated chemical potential $\mu $ for the
variational calculation (Eqs. (\ref{b}) and (\ref{Var})), the solution
following Eq. (\ref{matriz}), soliton solution (\ref{soExa}), and the
numerical implementation for the GPE. The numerical procedure for the
calculation of $\mu /\hbar \omega $ was implemented in order to achieve a
maximum uncertainty of $10^{-10}$. According to the results of Fig. 7, the
system (\ref{Hill}) using 50$\times $50 matrix reproduces quite well the
chemical potential values in the interval $\Lambda >-10$ with an accuracy
less than 1.2\%, while for $\mu _{s},$ given by (\ref{soExa}), the relative
error tends to zero as $\Lambda $ decreases. The best accuracy for the
solution (\ref{Var}) is reached in the interval $-3<\Lambda <0$ and fails
for smaller values of $\Lambda .$ In the inset, we show the\ calculated
chemical potential in the framework of a perturbation method, Eq. (\ref{pert}%
), and compared with the other four methods.\ Here, it can be seen the
strong deviation of the soliton solution from the correct values for $%
\Lambda >-2.$ However, no differences are observed between the numerical,
perturbation method, and the calculations using (\ref{Hill}). Again as in
the repulsive case, the results shown in Figs. 6 and 7 are of universal
validity, giving an absolute estimation of the accuracy of each employed
method as a function of a unique dimensionless parameter $\Lambda .$ The
present results teach us the way to get simple and exact analytical
solutions for the GPE in the attractive interaction case in terms of $%
\Lambda .$ Indeed, for $\Lambda \gtrsim -10$ using a small base (of the
order of 50 oscillator wave functions) we obtain an accuracy of $10^{-8}$
for the chemical potential along with a minimum accumulated error, $\eta ,$
of 0.005 for the order parameter. For smaller values of $\Lambda $ the
soliton solution (\ref{so}) and (\ref{soExa}) can be implemented as the
exact solution of Eq. (\ref{GP1}). At this point, it is necessary to analyze
the convergence to the exact solution provided by the series (\ref{fi}). In
principle, as it was derived in Sec. II, the function (\ref{fi}) is an exact
representation of the order parameter $\Phi $ with a convergence at list in
energy to the real order parameter $\Phi .$ The basis \{$\varphi _{n}(x)$\}
is a complete set for the Hilbert space defined by Eq. (\ref{GP1})
independent of the sign of the non-linear interaction term. Nevertheless,
the number of the harmonic oscillator wave functions needed to reach the
necessary convergence to the real solution depends on the values and sign of
$\Lambda $. The key point is to know when the series (\ref{fi}) is really a
good method for calculations and more efficient than the numerical ones. In
our case, we selected 50 even functions $\varphi _{n}$ reaching an accuracy
for the chemical potential less than $10^{-8}$ in the range $-10<\Lambda <25.
$ To get the same accuracy for the chemical potential in the attractive
region with $\Lambda <-10,$ it is necessary to deal with matrixes (\ref%
{matriz}) of rank larger than 50$\times $50.

In order to clarify this peculiarity of the expansion (\ref{fi}) we define
the effective potential

\begin{equation*}
Veff=\frac{1}{2}m\omega ^{2}x^{2}+\lambda \left\vert \Phi _{num}\right\vert
^{2},
\end{equation*}%
where the order parameter $\Phi $ has been substituted by the numerical
solution $\Phi _{num}.$ Figure 8 shows the potential $Veff$ in units of $%
\hbar \omega $ for both, the attractive and repulsive interactions. In the
figure, we represented the exact calculation of $\mu $ for each considered
value of $\Lambda $. It becomes clear that for the repulsive case, $Veff$
resembles the harmonic oscillator potential (Fig. 8 a)) and the chemical
potential falls within certain range of the harmonic oscillator eigenvalues.
Hence, the complete set of harmonic wave function \{$\varphi _{n}(x)$\} can
reproduce well, with an inexpensive computational effort, the mathematical
properties of the GPE. In the case of attractive interaction, see Fig. 8 b),
the situation changes drastically. Here, the effective potential becomes
more localized as $\Lambda $ decreases and for $\Lambda \rightarrow -\infty
, $ $Veff\sim \delta (x).$ The function $Veff$ does not resemble the
harmonic oscillator potential, thus the values of the chemical potential are
far away from $(n+%
{\frac12}%
)$ eigenvalues. Although the basis \{$\varphi _{n}(x)$\} is complete, the
number of functions $\varphi _{n}(x)$ needed to describe the order parameter
$\Phi $ and chemical potential $\mu $ with certain accuracy should increase
enormously as $\Lambda $ decreases. This performance of the attractive
interaction, determines that the Green function solution or equivalently the
Galerkin or spectral method becomes computational expensive and the method
is not adequate to describe the GPE for strong attractive interaction case,
that is for $\Lambda <-10$.

\section{Conclusions}

We have provided simple analytical forms to get explicit solutions for the
GPE. The reported analytical techniques allow us to explore regions of
positive and negative nonlinear interactions in condensates. We estimated
the range of applicability of the perturbation theory, Thomas-Fermi
approximation, soliton wave function, soliton variational calculation, and
Green function solution (spectral method) through a universal interaction
parameter $\Lambda =\lambda /l_{0}\hbar \omega .$ The perturbation method is
valid in the weak interaction limit, $-2$ $l_{0}\hbar \omega <\lambda <2$ $%
l_{0}\hbar \omega $ with an error for the chemical potential less than 1.5\%
while the TF approximation provides an error less than 3\% if $\lambda
\eqslantgtr 10$ $l_{0}\hbar \omega .$ The solution (\ref{fi}) with solely 50
harmonic oscillator wave functions reproduces quite well the chemical
potential $\mu $ with an accuracy of 1.2\% in the interval $-10$ $l_{0}\hbar
\omega <\lambda <10$ $l_{0}\hbar \omega .$ We identified that the series (%
\ref{fi}) or the spectral method is not adequate and can be computational
expensive for the attractive case if $\lambda <-10$ $l_{0}\hbar \omega $
(see Figs. 6, 7 and 8). In this case, the bright soliton solutions (\ref{so}%
) and (\ref{soExa}) represent the better approach for the order parameter
and the chemical potential respectively$.$ The presented soliton limit is
formally equivalent to the Thomas-Fermi one and becomes a powerful tool for
condensates with strong attractive interaction. Also, we have introduced a
soliton variational procedure valid for repulsive and attractive
interactions which can be applied to the study of the dynamics of BEC or to
model physical systems obeying the GPE. With the present results it is
possible to have a short and comprehensive discussion on the usefulness of
different approaches for the mathematical and physical description of the
BEC.

We should note that the mathematical models here developed can be
straightforward extended to the three-dimensional case,\cite{pitaevskii}
two-dimensional "pancake-shaped",\cite{salasnich} or to the "cigar-shaped"
BEC's\cite{1D1,1D2,1D3} and to study the dynamics of two component BEC
systems.\cite{DobleC}

\acknowledgments

This work was supported in part from the Red de Macrouniversidades P\'{u}%
blicas de America Latina Exchange Program, from the Science Division of the
The City College of CUNY and from the CUNY-Caribbean Exchange Program. J. C.
D-P. is grateful to UFSCar and FFCLRP-USP\ for hospitality. V. L-R
acknowledge the financial support from Brazilian agencies FAPESP\ and CNPq.

\appendix

\section{Variational calculation}

Inserting the wave function (\ref{so}) in Eq. (\ref{GP1}) follows the Eq. (%
\ref{chva}), where $\alpha ,$ $\beta ,$ and $\gamma $ are numbers equal to:

\begin{eqnarray}
\gamma &=&\int_{-\infty }^{\infty }\sec h^{4}zdz=\frac{4}{3},  \label{u^2} \\
\alpha &=&2\gamma -\int_{-\infty }^{\infty }\sec h^{2}zdz=\frac{2}{3}, \\
\beta &=&\int_{-\infty }^{\infty }z^{2}\sec h^{2}zdz=\frac{\pi ^{2}}{6}
\end{eqnarray}

\section{Matrix elements}

The fourth dimensional matrix $\mathbf{T}$ introduced in Eq. (\ref{matriele}%
) is defined as

\begin{eqnarray}
T_{plmn} &=&\frac{1}{\pi \sqrt{2^{n+m+l+p}n!m!l!p!}}\times  \notag \\
&&\int \exp (-2z^{2})H_{n}(z)H_{m}(z)H_{l}(z)H_{p}(z)dz.  \notag \\
&&
\end{eqnarray}%
The matrix elements $T_{plmn}$ have the followings properties:

i) $T_{plmn}=0$ if $n+m+l+p=$odd number.

ii) $T_{plmn}$ is invariant under the permutation of the quantum numbers $n$%
, $m$, $l$, and $p$, i.e. $T_{plnm}=T_{lpmn}=T_{pmln}=...$

iii) For $m=0$ we find\cite{Gradshteyn80}

\begin{equation}
T_{pln0}=\frac{2^{s-1}}{\pi ^{2}}\frac{\Gamma (s-l)\Gamma (s-p)\Gamma (s-n)}{%
\sqrt{2^{n+l+p}l!p!n!}},
\end{equation}%
where $\Gamma (z)$ is the gamma function and $2s=n+l+p+1.$

iv) The following relations hold between two successive matrix elements $%
T_{pln0}:$

\begin{equation}
T_{pln0}=\frac{(s-l-1)(s-n-1)}{(s-p)\sqrt{p(p-1)}}T_{p-2ln0},
\end{equation}%
or

\begin{equation}
T_{pln0}=\frac{(s-n-1)}{\sqrt{lp}}T_{p-1l-1n0},
\end{equation}%
with

\begin{equation}
T_{0000}=\frac{1}{\sqrt{2\pi }}.
\end{equation}

v) For the most general case we have the expression\cite{lord}%
\begin{eqnarray}
T_{p,l,n,m} &=&\frac{(-1)^{M-m-p}2^{M-\frac{1}{2}}}{\pi \sqrt{%
2^{n+m+l+p}n!m!l!p!}}\times  \notag \\
&&\frac{\Gamma (M-l+\frac{1}{2})\Gamma (M-n+\frac{1}{2})}{\Gamma (M-n-l+%
\frac{1}{2})}\times  \notag \\
&&\text{ }_{\text{3}}F_{\text{2}}\left(
\begin{array}{cc}
-m,\text{ \ }-p, & -M+n+l+\frac{1}{2}; \\
-M+l+\frac{1}{2}, & -M+n+\frac{1}{2};%
\end{array}%
1\right) ,  \notag \\
&&
\end{eqnarray}%
where $_{\text{3}}F_{\text{2}}\left(
\begin{array}{ccc}
\alpha _{1}, & \alpha _{2}, & \alpha _{3}; \\
& \beta _{1}, & \beta _{2};%
\end{array}%
1\right) $ is the generalized hypergeometric series\cite{Gradshteyn80} and $%
2M=p+l+m+n.$

vi) The matrix element $T_{plnm}$ satisfies the recurrence relation

\begin{eqnarray}
T_{plnm} &=&\sqrt{\frac{n+1}{m}}T_{pln+1m-1}-  \notag \\
&&\sqrt{\frac{l}{m}}T_{pl-1nm-1}-\sqrt{\frac{p}{m}}T_{p-1lnm-1}.
\end{eqnarray}%
These mathematical properties allow to evaluate the tensor $\mathbf{T}$ in a
straightforward way and in consequence to solve Eq. (\ref{matriz}) for the
eigenvalues $\mu $ and eigenvector $\mathbf{C}$ very efficiently.


\begin{thebibliography}{99}
\bibitem{Exp1} M. H. Anderson, J. R. Ensher, M. R. Matthews, C. E. Wieman,
E. A Cornell, Science \textbf{269,} 198 (1995).

\bibitem{Exp2} K. B. Davis, M. -O. Mewes, M. R. Andrews, N. J. van Druten,
D. S. Durfee, D. M. Kurn, and W. Ketterle, Phys. Rev.Lett. \textbf{75,} 3969
(1995).

\bibitem{Exp3} C. C. Bradley, C. A. Sackett, J. J. Tollett, and R. G. Hulet,
Phys. Rev. Lett. \textbf{75,} 1687 (1995).

\bibitem{Exp4} D. J. Han, R. H. Wynar, Ph. Courteille, and D. J. Heinzen,
Phys. Rev. A \textbf{57}, 4114 (1998).

\bibitem{Exp5} T. Esslinger, I. Bloch, and T. W. H\"{a}nsch, Phys. Rev. A
\textbf{58,} 2664 (1998).

\bibitem{Exp6} L. V. Hau, B. D. Busch, Ch. Liu, Z. Dutton, M. M. Burns, and
J. A. Golovchenko, Phys. Rev. A \textbf{58,} 54 (1998).

\bibitem{Exp7} C. C. Bradley, C. A. Sackett, J. and R. G. Hulet, Phys.
Rev.Lett. \textbf{78,} 985 (1997).

\bibitem{Groos1} E. P. Gross, Nuovo Cimento \textbf{20} 454 (1961); L. P.
Pitaevskii, Zh. Eksp. Teor. Fiz. \textbf{40} 646 (1961) [1961 Sov. Phys.
JETP 13 451].

\bibitem{pitaevskii} F. Dalfovo, S. Giorgini, L. P. Pitaevskii, and S.
Stringari, Rev. Mod. Phys. \textbf{71}, 463 (1999).

\bibitem{khaykovich} L. Khaykovich, F. Schreck, G. Ferrari, T. Bourdel, J.
Cubizolles, L. D. Carr, Y. Castin, C. Salomon., Science \textbf{296}, 1290
(2002); K.E. Strecker, G. B. Partridge1, A. G. Truscott and R. G. Hulet1.,
Nature (London) \textbf{417}, 150.(2002).

\bibitem{burger} S. Burger, K. Bongs, S. Dettmer, W. Ertmer, and K.
Sengstock, Phys. Rev. Lett. \textbf{83}, 5198 (1999).

\bibitem{Eiermann} B. Eiermann, Th. Anker, M. Albiez, M. Taglieber, P.
Treutlein, K.-P. Marzlin, and M. K. Oberthaler, Phys. Rev. Lett. \textbf{92}%
, 230401 (2004).

\bibitem{Bogoliuv} L. P. Pitaevskii, Zh. Eksp. Teor. Fiz. \textbf{40}, 646
(1961) [JETP \textbf{13}, 451, (1961)].

\bibitem{ruprecht} P. A. Ruprecht, M. Edwards, K. Burnett, and C. W. Clark,
Phys. Rev. A \textbf{54,} 4178 (1996).

\bibitem{Pomoraev} A. V. Ponomarev, J. Madro\~{n}ero, A. R. Kolovsky, and A.
Buchleitner. Phys. Rev. Lett. \textbf{96}, 050404 (2006).

\bibitem{1D1} V. M. P\'{e}rez-Garc\'{\i}a, H. Michinel, and H. Herrero,
Phys. Rev. A \textbf{57}, 3837 (1998).

\bibitem{trallero} C. Trallero-Giner, J. Drake, V. Lopez-Richard, C.
Trallero-Herrero, Joseph L. Birmand, Physics Letters A \textbf{354}, 115
(2006).

\bibitem{edwards} M. Edwards and K. Burnett, Phys. Rev. A \textbf{51}, 1382
(1995).

\bibitem{baym} G. Baym and C. J. Pethick, Phys. Rev. Lett. \textbf{76}, 6
(1996).

\bibitem{morse} P. M. Morse and H. Feshbach, \textit{Methods of Theoretical
Physics} (NY, McGraw-Hill, 1953).

\bibitem{Mihling} S. G. Mikhlin and K. L. Pr\"{o}ssdorf, \textit{Approximate
Methods for Solutions of Differential and Integral Equations (}American
Elsevier Publ. Co., NY, 1967).

\bibitem{petrovskii} I. G. Petrovskii, \textit{Lectures on the Theory of
Integral Equaions} (Graylock Press, Rochester, 1957).

\bibitem{Abramowitz} \emph{Handbook of Mathematical Functions}, edited by M.
Abramowitz and I. Stegun (Dover, NY, 1972).

\bibitem{MihlingII} S. G. Mikhlin, \textit{Variational Methods in
Mathematical Physics} (Pergamon Press, 1964).

\bibitem{Hill} Bender C M and Orszag S A 1978 \textit{Advanced Mathematical
Methods for Scientists and Engineers} (NY, Mc Graw-Hill).

\bibitem{Pu} H. Pu and N. P. Bigelow, Phys. Rev. Lett. \textbf{80}, 1130
(1998).

\bibitem{salasnich} L. Salasnich, A. Parola, and L. Reatto, Phys. Rev. A
\textbf{65}, 043614 (2002).

\bibitem{1D2} Yu. S. Kivshar, T. J. Alexander, and S. K. Turitsyn, Phys.
Lett. A \textbf{278}, 225 (2001).

\bibitem{1D3} F. Kh. Abdullaev et al., Phys. Rev. Lett. \textbf{90}, 230402
(2003).

\bibitem{DobleC} C. J. Myatt, E. A. Burt, R. W. Ghrist, E. A. Cornell, and
C. E. Wieman, Phys. Rev. Lett. \textbf{78}, 586 (1997).

\bibitem{Gradshteyn80} I. S. Gradshteyn and I. M. Ryzhik\textit{, Tables of
Integrals, Series and Products }(Academic, NY, 1980)\textit{.}

\bibitem{lord} R. D. Lord, J. London Math. Soc., \textbf{24}, 101 (1949).
\end{thebibliography}
\end{document}